\documentclass[prb,twocolumn,preprintnumbers,amsmath,amssymb,showpacs,showkeys,superscriptaddress]{revtex4}

\usepackage[usenames]{color}
\usepackage{graphicx}
\usepackage{amsfonts}
\usepackage{amssymb}
\usepackage{epsfig}
\usepackage{bm}
\usepackage{dcolumn}
\newcommand{\beq}{\begin{eqnarray}}
\newcommand{\eeq}{\end{eqnarray}}
\newcommand{\e}{{\text e}}

\newcommand{\mf}{\textrm{{\scriptsize MF}}}
\newcommand{\imf}{\textrm{{\scriptsize CMF}}}

\def\bk{k}

\def\GS{0}
\def\es{{\textrm{e.s.}}}
\def\os{{\textrm{o.s.}}}

\begin{document}
\title{Phase diagram of the hardcore Bose-Hubbard model on a checkerboard superlattice}
\date{\today}

\author{Itay Hen}
\email{itayhe@physics.georgetown.edu}
\affiliation{Department of Physics, Georgetown University, Washington, DC 20057, USA}
\author{M. Iskin}
\email{miskin@ku.edu.tr}
\affiliation{Department of Physics, Ko\c c University, Rumelifeneri Yolu, 34450 Sariyer, Istanbul, Turkey}
\author{Marcos Rigol}
\email{mrigol@physics.georgetown.edu}
\affiliation{Department of Physics, Georgetown University, Washington, DC 20057, USA}

\begin{abstract}
We obtain the complete phase diagram of the hardcore Bose-Hubbard model in the presence of a 
period-two superlattice in two and three dimensions. First we acquire the phase 
boundaries between the superfluid phase and the `trivial' insulating phases of 
the model (the completely-empty and completely-filled lattices) analytically. Next, the boundary between 
the superfluid phase and the half-filled Mott-insulating phase is obtained numerically, 
using the stochastic series expansion (SSE) algorithm followed by finite-size scaling.
We also compare our numerical results against the predictions of several approximation schemes, 
including two mean-field approaches and a fourth-order strong-coupling expansion (SCE),  
where we show that the latter method in particular is successful in producing 
an accurate picture of the phase diagram. Finally, we examine the extent to which 
several approximation schemes, such as the random phase approximation 
and the strong-coupling expansion, give an accurate description of the momentum 
distribution of the bosons inside the insulating phases. 
\end{abstract}

\maketitle

\section{Introduction}
One of the most remarkable achievements in the field of ultracold Bose gases in recent 
years has been the observation of a superfluid to Mott-insulator transition in optical 
lattices.\cite{greiner02} By playing with the intensity of the different laser beams 
involved in the setup, experimentalists have been able to study this transition 
in effective one,\cite{stoferle04} two,\cite{spielman0708} and three\cite{greiner02} 
dimensional geometries. This extraordinary accomplishment was achieved with gases of 
bosonic atoms confined in optical and magnetic traps. Using the strength of the optical lattice 
as a control parameter, these gases were reversibly tuned from a Bose-Einstein condensate 
to a Mott insulator (a state composed of localized atoms) 
.\cite{bloch08}

It is generally accepted that this quantum phase transition can be studied using the 
Bose-Hubbard model, where the transition is found to be from a compressible superfluid 
phase to an incompressible Mott-insulating one (SF-MI).\cite{fisher} Over the years, 
much theoretical work has been devoted to determining the phase diagram of the model 
in various dimensions, using many different 
approaches.\cite{fisher,batrouni90,batrouni92,freericks94,freericks96,kuhner98,prokofiev07,prokofiev08} 
However, a direct comparison between theoretical results and experimental ones\cite{greiner02,stoferle04,spielman0708} still remains obscured by issues such 
as the spatial inhomogeneity,\cite{batrouni02,wessel04,rigol09} finite-temperature 
effects,\cite{ho07,gerbier07} and the limited set of experimental tools available 
to probe the nearly isolated ultracold atomic systems.

In a recent paper, Aizenman {\it et al.}\cite{Aizenman} argued that the phases of 
the Bose-Hubbard model can be studied equally-well by examining a slightly different 
variant of it, namely the Bose-Hubbard model in the limit of infinite onsite 
repulsion (i.e., the case of hardcore bosons), in the presence of an alternating (checkerboard)
onsite chemical potential (a superlattice with period two). The advantage of studying
the latter model lies in the fact that it exhibits all the salient properties of the 
Bose-Hubbard model, while also being more amenable to analytical treatment. Specifically, 
Aizenman {\it et al.} rigorously proved the existence of SF and 
MI phases in the half-filled three-dimensional case (although they did not
show that there is no intermediate phase between the two). In Ref.\ \onlinecite{hen09}, two of us (I.H. and M.R.) 
studied that very same model for the case of zero chemical potential 
both in two and three dimensions, using quantum 
Monte Carlo simulations and analytical approximation approaches. 
We showed that the SF-MI phase transition is a direct transition, and we determined its critical value.

The hardcore Bose-Hubbard model with a superlattice has yet another attractive feature
that the general Bose-Hubbard model lacks: it is exactly solvable 
in one dimension. This is due to the existence of a mapping of the hardcore bosons to noninteracting fermions.
This in turn enables the evaluation of correlation functions of interest 
by exact means.\cite{Val,rigol04,rigol06}

In this paper, we study the complete phase diagram of hardcore bosons in the presence of a superlattice 
in two and three dimensions and with arbitrary chemical potential. We determine the phase 
boundaries separating the compressible SF phase of the model from the various 
insulating phases. First we acquire the phase boundaries between the 
SF phase and the `trivial' insulating phases (the completely-empty and 
completely-filled lattices) analytically. 
Then we perform high-precision numerical simulations using the 
stochastic series expansion (SSE) algorithm\cite{SSE1,SSE2} in order to find the phase 
boundary of the transition between the SF and the half-filled MI. 
This is done by calculating the free energy $\Omega$, the density of bosons in the 
zero-momentum mode $\rho_0$, and the superfluid density $\rho_s$. The latter two 
quantities drop to zero upon entering the insulating regime from the SF phase. 

Once the complete phase diagram is obtained, we proceed to examine the model analytically 
by employing two mean-field-type approximations and a strong-coupling perturbation 
scheme (up to fourth order in the hopping parameter) in order to determine the extent 
to which analytical methods allow a reliable description of the system and its various 
physical properties, specifically in the context of the phase boundaries separating 
the compressible SF regime from the incompressible insulating regions. 

The paper is organized as follows. In Sec. \ref{sec:ibh} we review the model 
at hand and present a qualitative description of its expected phase diagram. In 
Sec.\ \ref{sec:ana}, we compute the phase boundaries between the SF phase 
and the empty and filled lattices analytically. 
In Sec.\ \ref{sec:SSE}, we obtain the remaining boundary between the SF
and the half-filled MI phase. This phase boundary is computed 
numerically, using the stochastic series expansion (SSE) algorithm. 
Section \ref{sec:MFSW} is devoted to studying the phase diagram as it is given 
by two mean-field approaches, and in Sec. \ref{sec:SCE} we employ a strong-coupling 
expansion (SCE) method. These approximation methods are then compared against the 
previously obtained numerically-exact results. In Sec. \ref{sec:md},
we study the momentum distribution of the bosons, in order to allow for
a comparison with future experimental data. Finally, in Sec.\ \ref{sec:conc}, we conclude with a 
discussion and summary of our results.

\section{\label{sec:ibh}Model}
The Hamiltonian for hardcore bosons in a period-two hypercubic superlattice in 
$d$-dimensions, with $N=L^d$ sites and periodic boundary conditions, can be written as:
\beq
 \label{eq:Ham}
\hat{H} = - t \sum_{\langle ij \rangle} \left( \hat{a}_i^{\dagger} \hat{a}_j 
+ \hat{a}_j^{\dagger} \hat{a}_i \right) - A \sum_i (-1)^{\sigma(i)} \hat{n}_i 
-\mu \sum_i \hat{n}_i \,. \nonumber\\
\eeq
Here, $\langle ij \rangle$ denotes nearest neighbors, $\hat{a}_i$ ($\hat{a}_i^{\dagger}$) 
destroys (creates) a hardcore boson on site $i$, $\hat{n}_i=\hat{a}_i^{\dagger} \hat{a}_i$ 
is the local density operator, $\mu$ is the global chemical potential, and 
$A(-1)^{\sigma(i)}$ is a checkerboard local potential with $\sigma(i)=0$ on the even 
sublattice and $1$ on the odd sublattice. The hopping parameter $t$ (which we shall fix at 
$t = 1$) sets the energy scale, and without loss of generality we choose $A > 0$.
The hardcore boson creation and annihilation operators satisfy the constraints
$\hat{a}^{\dagger 2}_{i}= \hat{a}^2_{i}=0$ and
$\lbrace  \hat{a}_{i},\hat{a}^{\dagger}_{i} \rbrace =1$, 
which prohibit double or higher occupancy of lattice sites, as dictated by the 
$U\rightarrow \infty$ limit of the Bose-Hubbard model. For any two different sites 
$i \neq j$, the creation and annihilation operators obey the usual bosonic relations
$[\hat{a}_{i},\hat{a}_{j}]=[\hat{a}^{\dagger}_{i},\hat{a}^{\dagger}_{j}]=
[\hat{a}_{i},\hat{a}^{\dagger}_{j}]=0$.

To understand the zero-temperature phase diagram of hardcore bosons in a superlattice
potential, let us first analyze the atomic ($t = 0$) limit. In this limit, there is no 
kinetic (hopping) term, and the boson number operators $\hat{n}_i$ commute with the 
Hamiltonian, so every lattice site is occupied by a fixed number of bosons. The average 
boson occupancy is determined so as to minimize the ground-state (free) energy. In 
particular, for $A = 0$, the model is translationally invariant, and the ground-state 
boson occupancy is the same for each of the lattice sites: for $\mu < 0$ the 
minimal energy configuration is simply the particle vacuum (VP),
i.e., the completely-empty lattice, and for $\mu > 0$ the minimal energy 
configuration is simply the hole vacuum (VH), i.e., the completely-filled lattice.
The ground-state energy of these phases is degenerate at $\mu = 0$. 
When $A \ne 0$, the ground state has an additional half-filled insulating phase 
characterized by crystalline order in the form of staggered boson densities, i.e.,
$\langle \hat{n}_i \rangle  = 1$ for the even (or odd, depending on the sign of $\mu/A$) sublattice and 
$\langle \hat{n}_i \rangle = 0$ for the odd (or even) one. 
We call this alternating density pattern the MI phase, 
although it is sometimes referred to as a charge density wave.\cite{Val}
The MI phase resides in the region $|\mu/A| <1$, sandwiched between 
the particle vacuum and the hole vacuum. 

\begin{figure}[htp!]
\includegraphics[angle=0,scale=1,width=0.48\textwidth]{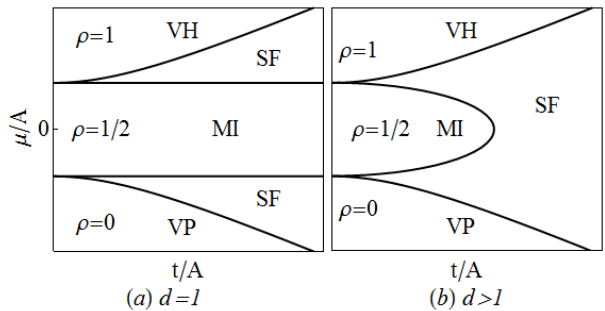}
\caption{\label{fig:lobe} Phase diagram of the hardcore Bose-Hubbard model in the presence 
of a period-two superlattice, Eq.\ (\ref{eq:Ham}). In one dimension (left panel), the phase 
diagram contains three incompressible insulating phases, indicated by VH -- the hole vacuum, 
i.e., a completely-filled lattice, VP -- the particle vacuum, i.e., the completely-empty lattice,
and MI -- the Mott insulator, in which case the average density is $1/2$ and the local 
densities on the even and odd sublattices are different. Outside of these insulating regions, 
the system is superfluid (SF). In higher dimensions (right panel), the phase diagram is 
similar, with one exception: while in one dimension the MI phase extends to infinity, 
in higher dimensions the MI phase takes the form of a Mott lobe.}
\end{figure}

Having discussed the $t = 0$ limit, we are now ready to analyze the competition 
between the kinetic and the potential energy terms of the Hamiltonian when 
$t \ne 0$. In one dimension, the phase diagram of the model is already known.
As noted in the Introduction, the model in this case has an analytic 
solution.\cite{Val,rigol06} This is due to the Jordan-Wigner transformation 
which enables the mapping of the hardcore boson Hamiltonian to that of 
noninteracting spinless fermions. The dispersion relation in this case is given by
\beq
\varepsilon(k)=-\mu \pm \sqrt{4 t^2 \cos^2(ka) +A^2} \,,
\eeq
where $a$ is the lattice constant. The phase diagram consists of three insulating 
incompressible regions (these are extensions of the $t=0$ ones), as shown in 
Fig.\ \ref{fig:lobe}(a). Two are the `trivial' insulators: the VP phase 
which is obtained for large and negative values of $\mu$, and the VH phase 
which is obtained for large and positive values of $\mu$. 
These two phases are also present in the absence 
of the alternating potential, and are particle-hole `mirror images' of each other. 
They are separated from the SF phase along the curves
$\mu/A = \pm \sqrt{1+(2t/A)^2}$ [see Fig.\ \ref{fig:lobe}(a)].
As evident from the expression for the dispersion relations given above, the superlattice 
(i.e., the onsite checkerboard potential) creates a gap of $\Delta=2A$ in the energy 
spectrum, leading to a MI phase at half filling. This is the `slab' enclosed 
by $\mu/A=1$ from above and $\mu/A=-1$ from below, in the center of the figure.

In dimensions higher than one [Fig.\ \ref{fig:lobe}(b)], the expected phase diagram 
of the hardcore Bose-Hubbard model is qualitatively similar to the one-dimensional 
case with one notable exception. Here, the MI region does not extend to infinity, 
but instead is a finite lobe, connecting the two SF regimes together. 

The phase diagram of the hardcore Bose-Hubbard model has one additional property 
resulting from it being invariant under the transformation 
$\hat{a}_i \to \hat{a}^{\dagger}_{i+\hat{r}}$ (where $\hat{r}$ denotes a shift 
of one lattice step in any of the possible directions). This symmetry operation,
which can be immediately read off from the Hamiltonian, corresponds to a particle-hole 
exchange combined with swapping the odd and even sublattices. It leads to a 
$\mu \to - \mu$ symmetry in the phase diagram. We shall make use of this fact when 
we obtain the phase diagram in later sections. The special case of $\mu=0$ has been 
studied in Ref.\ \onlinecite{hen09} both in two and three dimensions.

Before moving on, we recall that the model at hand can also be viewed 
as the $XY$ model of a spin-1/2 system.\cite{mats,lieb61} This is due to the 
mapping between bosonic operators and $SU(2)$ generators: 
\beq
\hat{a}_i^{\dagger} &\leftrightarrow & \hat{S}_i^{+} \,, \\\nonumber
\hat{a}_i &\leftrightarrow & \hat{S}_i^{-} \,, \\\nonumber
\hat{a}_i^{\dagger} \hat{a}_i &\leftrightarrow & \hat{S}_i^{z}+1/2 \,.
\eeq
With this mapping, the hardcore bosons Hamiltonian, \hbox{Eq.\ (\ref{eq:Ham})}, 
becomes that of the $XY$ antiferromagnet with an alternating magnetic field 
applied along the $\hat{z}$ direction:
\beq
\hat{H}= &-& t \sum_{\langle ij \rangle} \left( \hat{S}_i^{+} \hat{S}_j^{-} 
+ \hat{S}_j^{+} \hat{S}_i^{-} \right) 
\nonumber \\ &-& \sum_i \left[ \mu 
+A (-1)^{\sigma(i)} \right] \left( \hat{S}_i^{z}+\frac1{2} \right)  \,.
\eeq
This alternative representation will become handy in the next sections.  

\section{ \label{sec:ana} Vacuum of particles and holes phase boundaries}

As it turns out, the phase boundary separating the SF phase 
from the insulating VH phase (henceforth, the SF-VH boundary) can be easily 
obtained analytically for any given dimension. To see this, we will use the 
fact that our Hamiltonian commutes with the total-number-of-bosons operator 
$\hat{N}=\sum_i \hat{n}_i$. In spin language, this simply means that for any 
given set of parameters $\mu$, $A$ and $t$, the ground-state wave function 
will be a linear combination of product states each having the same number 
of spin-downs. In the VH phase, this number is zero, as the wave function 
is simply 
\beq \label{eq:vh}
|\textrm{VH} \rangle =| \uparrow\uparrow\uparrow\ldots\uparrow\uparrow\uparrow\rangle\,,
\eeq
with energy $\varepsilon_{\textrm{vh}} = -\mu N$. In the infinitesimally thin layer 
outside the VH phase, the state of the system (which we shall refer to as the VH 
`defect' state) is characterized by exactly one spin-down. That is, the wave function 
has the form:
\beq \label{eq:vhdef}
| \textrm{VH}_{\textrm{def}} \rangle = \sum_i c_i \hat{S}_i^{-} | \textrm{VH} \rangle \,.
\eeq
The symmetry of our model further tells us that all the coefficients $c_i$ whose index 
`$i$' corresponds to a site on the even (odd) sublattice are all the same, namely:
\beq
c_i=\frac{c_{\es}+c_{\os}}{2} + (-1)^{\sigma(i)} \frac{c_{\es}-c_{\os}}{2}\,,
\eeq
where normalization requires $N/2( |c_{\es}|^2 +|c_{\os}|^2)=1$, and $\es$ ($\os$) 
stands for the even (odd) sublattice. In order to determine the exact value of the weights 
$c_{\es}$ and $c_{\os}$, we first act with the Hamiltonian on this state. This eigenvalue 
problem then reduces to the following coupled equations:
\begin{subequations}
\begin{align}
-2 d t \, c_{\os} +[\mu(1-N) +A] c_{\es} &=\varepsilon \, c_{\es} \\
-2 d t \, c_{\es} +[\mu(1-N) -A] c_{\os} &=\varepsilon \, c_{\os}\,,
\end{align}
\end{subequations}
where $\varepsilon$ is the energy of the state. Solving for $\varepsilon$, the solution
with minimal energy turns out to be
\beq
\varepsilon_{\textrm{def}} = -\mu N +\mu -\sqrt{A^2 +(2 d t)^2} \,.
\eeq
The SF-VH boundary is the curve along which the VH state, Eq.\ (\ref{eq:vh}), is no 
longer energetically favorable. This happens when its energy becomes equal to the energy 
of the defect state, Eq.\ (\ref{eq:vhdef}). Matching the two, we obtain the SF-VH 
phase boundary:
\beq \label{eq:vhbranch}
\frac{\mu}{A} = \sqrt{1+x^2} \,,
\eeq
where $x = 2 d t /A$. 

A few remarks are now in order. As already noted in the previous section, the phase diagram 
of the hardcore Bose-Hubbard model is symmetric under the transformation $\mu \to -\mu$.
This tells us that the SF-VP phase boundary [the lowest branch in Fig.\ \ref{fig:lobe}(b)], 
is given by $\mu/A =-\sqrt{1+x^2}$. This result can also be obtained by repeating the 
above exercise with the substitution $| \uparrow \rangle \leftrightarrow | \downarrow \rangle$. 
We also note that Eq.\ (\ref{eq:vhbranch}) agrees with the corresponding 
expression of the one-dimensional case obtained formerly (see Sec.\ \ref{sec:ana}).

Another, simpler argument leading to the same solution stems from the fact that 
the boundary between the SF and the VP (VH) phase is determined 
by the addition of a single particle (hole) to the completely-empty (-filled) lattice.
It can then be argued that whether one is dealing with hardcore bosons or noninteracting 
spinless fermions makes no difference in this case, as the particle statistics plays no role. 
This further means that one needs only to diagonalize the single-particle Hamiltonian 
and find the energy difference between the completely-empty (-filled) lattice and the state with
one particle (hole). These will provide the chemical potential at the boundary between 
the SF and the VP (VH) phase. The single-particle spectrum in a $d$-dimensional 
superlattice with period two has the form:
\beq
\varepsilon(k)=-\mu \pm \sqrt{4d^2t^2 \cos^2(ka) +A^2} \,,
\eeq
from which Eq.\ (\ref{eq:vhbranch}) follows trivially.

\section{\label{sec:SSE} Numerical results}

Unlike the SF-VH and SF-VP phase boundaries, the SF-MI boundary, cannot be determined 
with the tools introduced in the previous section. One reason for that is that the 
exact many-body wave function of the MI state is not known. Therefore, 
in this section we explore the SF-MI phase boundary numerically by performing simulations 
based on the stochastic series expansion (SSE) algorithm.\cite{SSE1,SSE2} Our main 
objective here is to find the critical points of the SF-insulator transitions 
in the $\mu$-$A$ parameter space (without loss of generality we fix the hopping parameter at
$t=1$ and consider only $\mu>0$ and $A>0$). Critical points on the SF-MI boundary 
were typically obtained by first fixing the value of the parameter $A$,
and then performing the simulations for a range of values of $\mu$ and different system sizes. 
This procedure was then repeated for different values of $A$. 
In some cases, mainly near the tip of the Mott lobe, 
we repeated the above procedure by fixing the value of $\mu$ 
and performing simulations for a range of $A$ values and different system sizes.
This was done mainly to further verify the accuracy of the results, as the tip of the lobe is a multicritical point
and therefore requires more care.

Repeating the simulations with different system sizes, enables us to extrapolate the 
thermodynamic limit by correcting finite-size effects using scaling arguments in the 
vicinity of the phase transition: around the critical point, most physical quantities 
(which we denote here by $X$) scale according to the general rule:
\beq \label{eq:scaling}
X L^{\xi/\nu} = F(|\mu-\mu_{\textrm{c}}| L^{1/\nu}) \,,
\eeq
where $F$ is a universal scaling function, $\mu-\mu_{\textrm{c}}$ is the shifted control 
parameter ($\mu$ being the control parameter, and $\mu_{\textrm{c}}$ its critical value), 
$\nu$ is the correlation length critical exponent and $\xi$ is the critical exponent 
belonging to the observable $X$. The values of these exponents are determined by the 
universality class the transition belongs to. In a previous work,\cite{hen09} we studied the SF-MI 
transition at fixed (half-filled) density. This type of transition belongs to the ($d$+1) $XY$
universality class, similarly to the SF to MI transition of the 
Bose-Hubbard model at fixed integer density.\cite{fisher} Here, we compute the 
phase boundary between the SF and the (half-filled) MI phase
while changing the density, so the transition belongs to 
the mean-field universality class for which the correlation length and dynamical critical
exponents are $\nu=1/2$ and $z=2$ (again, exactly as the corresponding transition in the Bose-Hubbard 
model).\cite{fisher}

Equation (\ref{eq:scaling}) above will help us find the critical point, as it tells us that 
(a) the quantity  $X L^{\xi/\nu}$ should be independent of the size of the system at the phase 
transition, and (b) when plotting $X L^{\xi/\nu}$ against $|\mu-\mu_{\textrm{c}}| L^{1/\nu}$ 
the resulting curve should be independent of the system-size as well. The quantity we shall 
be using to that end is the superfluid density, which has the critical exponent 
$\xi=\nu(d+z-2)$ (see Ref.\ \onlinecite{fisher} for details) where $d$ is the dimension.

We note here that since we are interested in the zero-temperature properties of the system,
simulations are performed with high inverse-temperature $\beta=1/T$ (in our units, $k_B=1$),
where in most cases we will find it sufficient to have $\beta \geq 2 L$ in order to obtain 
virtually zero-temperature results. (The effects of increasing $\beta$ beyond this value are 
indiscernible.)

As already discussed, in one dimension, our model has an analytic solution.\cite{Val}
This enabled us to compare our numerical method against exact analytic results,
as a check on our computational approach. No discrepancies between the analytical 
solution and the numerical one were found (see also Ref.\ \onlinecite{hen09}). 

\begin{figure}[htp!]
\includegraphics[angle=0,scale=1,width=0.45\textwidth]{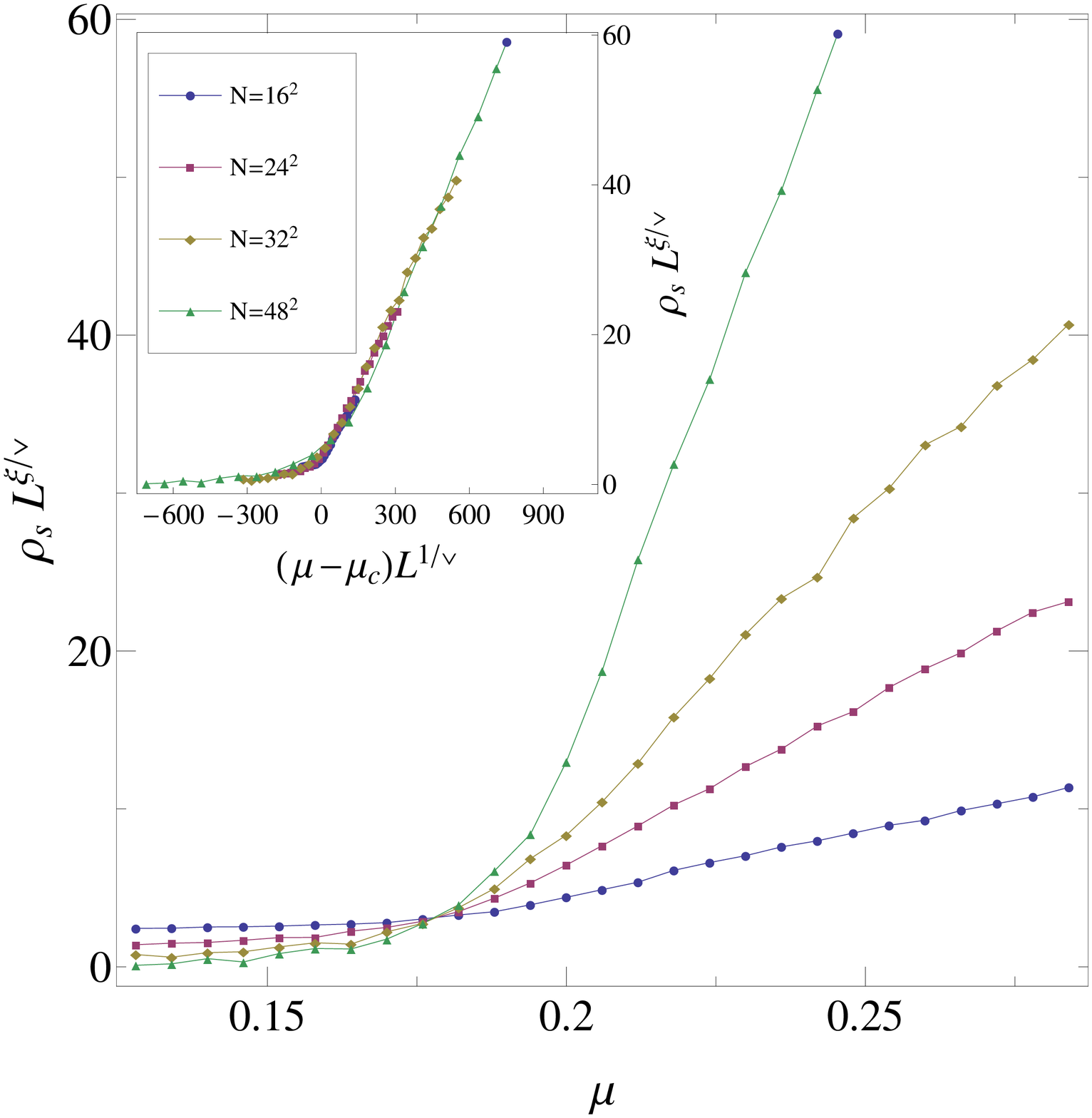}
\caption{\label{fig:fss2d} (Color online) Scaled superfluid density as a function of the 
chemical potential $\mu$ for the various system sizes in the two-dimensional case (here, 
$A=1.05$). All the curves intersect at $\mu \approx 0.178$ indicating the value of the 
critical point. In the inset, the control parameter (the horizontal axis) is scaled as well, 
leading to the collapse of all data points into a single curve.}
\end{figure}

\begin{figure}[htp!]
\includegraphics[angle=0,scale=1,width=0.49\textwidth]{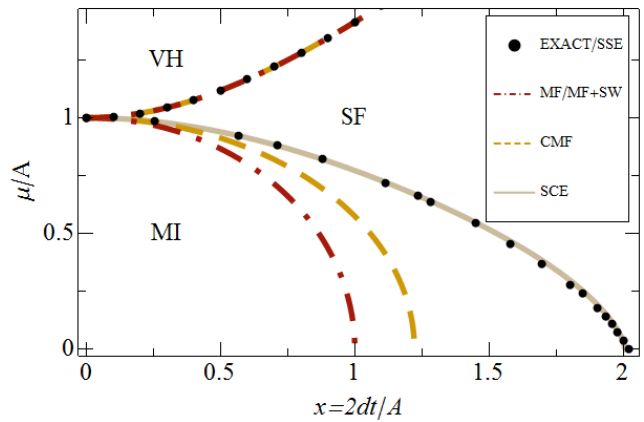}
\caption{\label{dim2} (Color online) Phase diagram of the model in two dimensions. The full 
circles are the analytical (VH boundary) and numerical (MI boundary) results. The solid 
line corresponds to the strong-coupling expansion (SCE) fit, whereas the dot-dashed and dashed 
lines are the mean-field (with and without spin-wave corrections) and cluster mean-field 
predictions, respectively. As the figure shows, the SF-VH boundary is predicted correctly 
by the mean-field approximation schemes. As for the SF-MI boundary, the predictions of the 
SCE fit provide the most accurate results.}
\end{figure}

In dimensions higher than one, no analytic solution to the model exists, so accurate results 
are obtainable only numerically. In the two dimensional case, we have applied the SSE algorithm 
to systems of sizes ranging from $16 \times 16$ to $48 \times 48$, with inverse-temperature 
$\beta=64$. Figure\ \ref{fig:fss2d} is an example of how scaling of the superfluid density 
data for the various system sizes is performed in order to find the critical point corresponding 
to $A=1.05$. Here, the scaled superfluid density is plotted against $\mu$ for the different 
system sizes (the statistical errors of the quantum Monte Carlo simulations are 
on the order of magnitude of the symbol sizes). All curves intersect at $\mu_{\textrm{c}} \approx 0.178$, 
signifying the phase transition for $A=1.05$. The inset shows the scaled superfluid 
density as a function of the scaled control parameter, in which case all curves should 
be, and in fact are, on top of each other. The resulting SF-MI phase boundary of our model 
in two-dimensions is marked by the full circles in Fig.\ \ref{dim2}. As noted earlier, 
the lower half of the phase diagram Fig.\ \ref{fig:lobe} (the $\mu<0$ half) is but a mirror 
image of the portion shown in Fig. \ref{dim2}, and thus is not presented there. 
The tip of the Mott lobe was found to be at $x_c \approx 2.02$.\cite{hen09} 

\begin{figure}[htp!]
\includegraphics[angle=0,scale=1,width=0.45\textwidth]{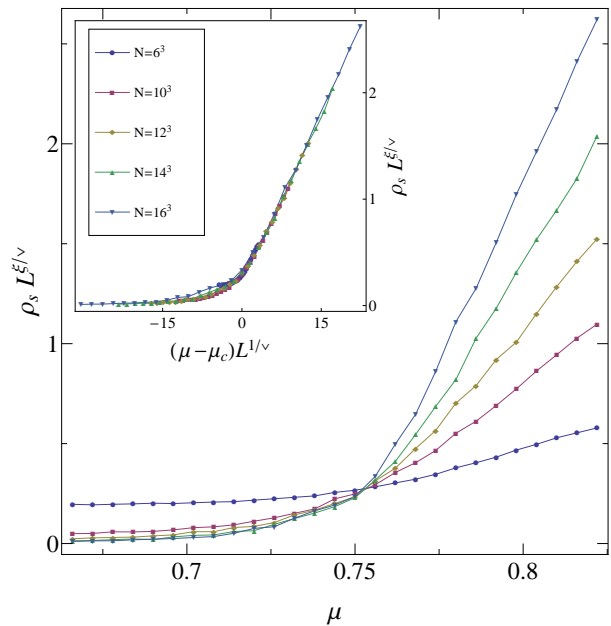}
\caption{\label{fig:fss3d} (Color online) Scaled superfluid density as a function of the 
chemical potential $\mu$ for the various system sizes in the three-dimensional case (here, 
$A=2.28$). All the curves intersect at $\mu \approx 0.752$ indicating the value of the 
critical point. In the inset, the control parameter (the horizontal axis) is scaled as well, 
leading to the collapse of all data points into a single curve.}
\end{figure}

\begin{figure}[htp!]
\includegraphics[angle=0,scale=1,width=0.49\textwidth]{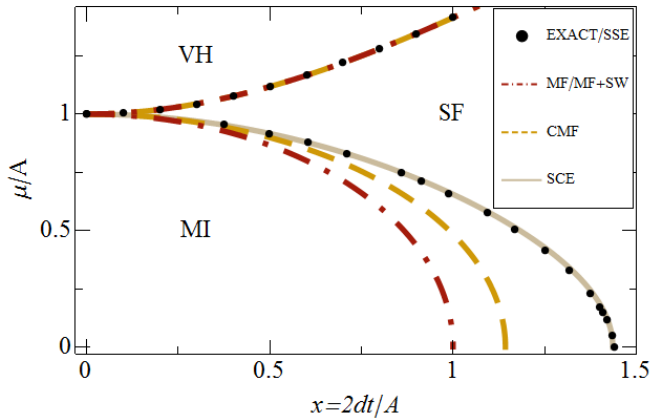}
\caption{\label{dim3} (Color online) Phase diagram of the model in three dimensions.
The full circles are the analytical (VH boundary) and numerical (MI boundary) results. 
The solid line corresponds to the strong-coupling expansion (SCE) fit, whereas the dot-dashed 
and dashed lines are the mean-field (with and without spin-wave corrections) and 
cluster mean-field predictions, respectively. As the figure shows, the SF-VH boundary 
is predicted correctly by the mean-field approximation schemes. As for the SF-MI boundary, 
the predictions of the SCE fit provide the most accurate results.}
\end{figure}

In three dimensions, we have performed simulations with system sizes ranging from 
$6 \times 6 \times 6$ to $16 \times 16 \times 16$ and an inverse temperature of $\beta=40$. 
Figure \ref{fig:fss3d} is an example of how scaling is carried out in three dimensions:
the scaled superfluid density is plotted as a function of $\mu$ for the different system 
sizes and $A=2.28$. The inset depicts the scaled superfluid density as a function of the 
scaled control parameter, exhibiting the collapse of all data points into a single curve, as 
in two dimensions. The resulting phase boundary in three-dimensions is shown in 
Fig.\ \ref{dim3} (full circles). The tip of the Mott lobe was found to be at $x_c \approx 1.44$.\cite{hen09}

\section{\label{sec:MFSW} Mean-field approaches}
Having obtained the exact boundaries of the phase diagram of the model, we now proceed to 
study several approximation schemes, and examine the extent to which they provide an 
accurate description of the phase diagram of the model. We start this investigation with 
the Gutzwiller mean-field approach. 

\subsection{\label{sec:mf} Gutzwiller mean-field}

Along the lines of Ref.\ \onlinecite{hen09}, we start our mean-field calculation with the 
following product state as our ansatz:
\beq \label{eq:GSMF}
| \GS \rangle_{\mf} =\prod_j^{\otimes} \left( \sin \frac{\theta_j}{2} | \downarrow \rangle + 
\cos \frac{\theta_j}{2} \e^{i \, \varphi_j} | \uparrow \rangle 
\right) \,.
\eeq
The angles $(\theta_j,\varphi_j)$ here, specify the orientation of the $j$-th spin.
Naturally, we expect the wave functions of each of the odd (even) sublattice sites to be
identical. This is due to the checkerboard symmetry of the model.

As we are using the grand-canonical scheme, the orientations of the spins  will be 
determined by minimizing the grand-canonical potential (per site)
\beq \label{eq:omegaMF}
\Omega_{\mf} &=& {_{\mf}}\langle \GS| \hat{H} | \GS\rangle_{\mf} 
=-\frac{t}{2 N} \sum_{\langle ij \rangle} \sin \theta_i \sin \theta_j \cos(\phi_i-\phi_j)
\nonumber\\
&\phantom{=}&-\frac1{2 N} \sum_i \left[ \mu +A (-1)^{\sigma(i)} \right] 
\left(1+\cos \theta_i\right)\,.
\eeq
with respect to these angles. For the azimuthal angles, this simply implies a constant (yet 
arbitrary) value \hbox{$\varphi_j=\Phi$}, while for the polar angles, the minimizers are
\begin{subequations} \label{eq:thetas}
\begin{align}
\cos \theta_1 &= \textrm{Min} \left[1,\textrm{Max} 
\left[-1, \mu_1 \sqrt{\frac{1+{\mu_2}^2}{1+{\mu_1}^2}} \right] \right]\,,\\
\cos \theta_2 &= \textrm{Min} \left[1,\textrm{Max} 
\left[-1, \mu_2 \sqrt{\frac{1+{\mu_1}^2}{1+{\mu_2}^2}} \right] \right]\,,
\end{align}
\end{subequations}
where $\mu_{1,2} = (\mu \pm A)/(2 d t)$. We note that while in 
Ref.\ \onlinecite{hen09} the focus was on the special $\mu=0$ case, here we
place no limitations on $\mu$. 

At this point we can calculate the following quantities. First, the average density 
of particles is:
\beq
\rho_{\mf} &=&\frac1{N} \sum_i {_\mf}\langle \GS | \hat{a}_i^{\dagger} \hat{a}_i | 
\GS \rangle_{\mf} = \frac1{2}+ \frac1{2 N}\sum_i \cos \theta_i
\nonumber\\
&=&\frac1{2} + \frac1{4} \left( \cos \theta_1 + \cos \theta_2\right)
\,.
\eeq
Next, the free energy becomes
\beq
\Omega_{\mf} &=& {_{\mf}}\langle \GS| \hat{H} | \GS\rangle_{\mf}
=- \frac{d t}{2} \sin \theta_1 \sin \theta_2 - \frac{\mu}{2} \nonumber\\
&-&\frac1{4} \left( \mu +A \right) \cos \theta_1
-\frac1{4} \left( \mu -A \right) \cos \theta_2 \,,
\eeq
and the density of bosons in the zero-momentum mode $\rho_0$ is calculated as:
\beq
\rho_{0,\mf} &=& \frac1{N} {_{\mf}}\langle \GS |\hat{a}^{\dagger}_{\bk=0} 
\hat{a}_{\bk=0}  | \GS \rangle_{\mf} \\\nonumber
&=&\frac1{4 N^2} \sum_{i,j} \sin \theta_i \sin \theta_j =\frac1{16} 
\left( \sin \theta_1 + \sin \theta_2 \right)^2 \,.
\eeq
\par
The superfluid density too is obtained in a straightforward manner.
In the mean-field approximation it has the simple form 
\hbox{$\rho_s = -(2d)^{-1} \partial \Omega / \partial t$}.\cite{hen09} 

\begin{figure}[htp!]
\includegraphics[angle=0,scale=1,width=0.49\textwidth]{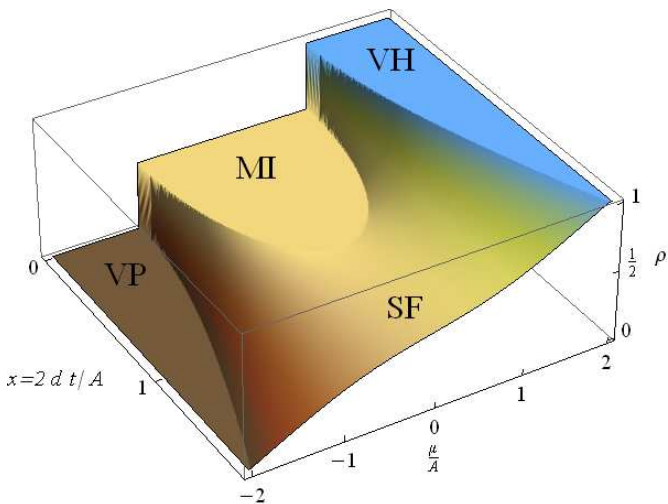}
\caption{\label{fig:mfPD} (Color online) Average density of bosons as a function of 
$x = 2 d t /A$ and $\mu/A$ in the mean-field approximation. The three insulating 
phases VP (empty lattice, zero density), MI (half-filled lattice) and 
VH (completely-filled lattice) are seen very clearly in the figure. 
Outside of these insulating regions is the SF phase.}
\end{figure}

The phase boundaries are simply the curves along which the superfluid density 
and the zero-momentum fraction drop to zero. These turn out to be:
\beq \label{eq:mfPM}
\frac{\mu}{A}=\sqrt{1 \pm x^2} \,, 
\eeq
where the `$+$' branch belongs to the SF-VH transition and the `--' branch belongs to 
the SF-MI transition (again, $x = 2 d t / A$). The phase diagram of the model as 
predicted by the Gutzwiller mean-field approach is sketched in Fig.\  \ref{fig:mfPD},
which shows the average density of bosons as a function of $x$ and $\mu/A$. 

\par

An alternative way of deriving the mean-field phase boundaries is through the decoupling 
approximation.\cite{fisher,stoof} In this approach, one approximates the hopping term as
\beq
\hat{a}_i^\dagger \hat{a}_j \approx \langle \hat{a}_i^\dagger \rangle \hat{a}_j
+ \hat{a}_i^\dagger \langle \hat{a}_j \rangle - \langle \hat{a}_i^\dagger \rangle  
\langle \hat{a}_j \rangle,
\eeq
and introduces the condensate order parameter $\psi_i = \sqrt{\hat{n}_i} = \langle 
\hat{a}_i^\dagger \rangle = \langle \hat{a}_i \rangle$ (analogous to the Bogoliubov 
approach). Since the condensate order parameter is the same for all lattice sites 
belonging to the same sublattice, i.e.,
\beq
\psi_i=\frac{\psi_{\es}+\psi_{\os}}{2} + (-1)^{\sigma(i)} \frac{\psi_{\es}-\psi_{\os}}{2}\,,
\eeq
for some real unknown parameters $\psi_{\es}$ and $\psi_{\os}$ (due to the checkerboard 
symmetry of the model), it is sufficient to solve only for the effective two-site Hamiltonian
\begin{align}
\label{eq:mfHam}
\hat{H}^{\rm MF}=
- 2dt \, \psi_{\es} (\hat{a}_j^{\dagger} + \hat{a}_j)
&- 2dt \, \psi_{\os} (\hat{a}_i^{\dagger} + \hat{a}_i) \nonumber \\
+ 4dt \, \psi_{\es} \psi_{\os} - A \hat{n}_i  + A \hat{n}_j &- \mu (\hat{n}_i + \hat{n}_j) \,,
\end{align}
where $i \in \es$ and $j \in \os$. 
Performing a second-order perturbation theory in the first two terms of this effective 
Hamiltonian around the VH and MI phases produces the ground state energies as a function of 
$\psi_{\es}$ and $\psi_{\textrm{o.s.}}$. Notice that higher orders are not needed for our 
purposes, since the second order theory is sufficient to derive the energy functional of the 
system up to second order in the order parameters $\psi_{\textrm{o.s.}}$ and 
$\psi_{\textrm{e.s.}}$. Following the usual Landau procedure for second-order phase 
transitions,  minimizing the ground state energies as a function of the superfluid order 
parameters, we eventually arrive at Eq.~(\ref{eq:mfPM}).

The dash-dotted lines in Figs.\ \ref{dim2} and \ref{dim3} show the phase diagram as 
predicted by the Gutzwiller mean-field approach, compared against the numerical 
results. Interestingly, the mean-field ansatz yields the correct solution for the SF-VH  
transition (upper branch). On the other hand, for the SF-MI boundary, mean-field results differ 
considerably from the numerical data: while away from the tip of the Mott lobe the method 
is very accurate, as one approaches the tip itself, errors climb up to their maximal values of 
$\approx 100\%$ in two-dimensions and $\approx 50\%$ in three dimensions at the tip of the 
MI lobe. The very large errors here reflect the fact that the mean-field approach is 
simply not fit to describe the phase transition in this region.

Before moving on, we remark here that addition of spin-wave corrections to the mean-field solution does not
modify the mean-field critical points of the model,\cite{hen09} so the phase boundary 
is not altered by spin-wave corrections. While deep in the SF phase spin-wave 
corrections yield major improvements over the mean-field results for many of our 
observables of interest, in terms of phase boundaries the spin-wave corrections do not 
contribute. As one approaches the phase transition itself, the spin-wave 
corrections lose their accuracy, eventually leaving the phase boundaries at their 
mean-field values.\cite{hen09} 

\subsection{Cluster mean-field}

Aiming to improve the results obtained in the previous section, we now describe a
`cluster' mean-field approach, which makes use of the 
checkerboard symmetry of the model. This approximation 
scheme was introduced in Ref.\ \onlinecite{hen09} where it was applied to the 
special case of $\mu=0$. Within this approach, one starts with a variational ansatz 
which, as before, is a product state. However, this time one does not choose a product 
of single-site wave functions. The new ansatz is a product of wave functions each 
describing the state of a `block' of $2^d$ sites, such that with this block as the basic 
cell, the model turns homogeneous. In two dimensions, for example, a block consists of 
$2 \times 2$ square cells each of which is described by the general wave function
\beq \label{eq:gsimf}
| \GS \rangle_{\imf} =\prod_{\textrm{blocks}}^{\otimes} 
\left(\sum_{i,j,k,l \in \{ \downarrow,\uparrow \}} c_{ijkl} | i j k l\rangle \right) \,,
\eeq
where the generalization to three dimensions, in which case the basic block is a 
$2 \times 2 \times 2$ cubic cell, is straightforward (note that the coefficients for 
each of the blocks will be the same due to the symmetry of the wave 
function).\cite{hen09} As before, we minimize the free energy 
\hbox{$\Omega_{\imf} = {_{\imf}}\langle \GS | \hat{H} | \GS \rangle_{\imf}$} 
with respect to the coefficients $c_{ijkl}$ of the wave function (this time we do 
so numerically). Obtaining the various observables in terms of the wave function 
given in Eq.\ (\ref{eq:gsimf}) is straightforward, and was performed in much the 
same way as the usual mean-field approach discussed in Sec.\ \ref{sec:mf}. 

The phase boundaries, as predicted by the cluster mean-field approximation, are given 
by the dashed lines in Figs.\ \ref{dim2} and \ref{dim3} for two and three dimensions, 
respectively. As the figures indicate, the SF-VH boundary is predicted correctly.
This is no surprise as the Gutzwiller mean-field, over which the current method 
is an improvement, is already exact for that boundary. As for the SF-MI boundary, the 
cluster mean-field method is far better than the Gutzwiller mean-field method. As in 
the previous mean-field case, the results are more accurate away from the tip of the 
Mott lobe but reach $\approx 60\%$ error in two dimensions $\approx 24\%$ error in 
three dimensions, as the tip is approached. 

Having shown that the mean-field-type theories presented here are not very accurate in 
describing the SF-MI phase boundary, in particular close to the tip of the lobe, we 
turn to develop a strong-coupling perturbation theory in the hopping $t$. This 
approach, combined with a scaling analysis, will allow us to predict the critical point 
and the shape of the insulating lobe in a more accurate manner.

\section{\label{sec:SCE} Strong-coupling expansion (SCE)}

Strong-coupling expansion (SCE) techniques were previously used to discuss the phase diagram 
of the Bose-Hubbard model,\cite{freericks94, freericks96,freericks09} and of the 
extended Bose-Hubbard model,\cite{iskin09} and its results showed an excellent agreement 
with quantum Monte Carlo simulations\cite{prokofiev07,prokofiev08} in the former case. 
Motivated by the success of this technique with Bose-Hubbard type models, here we 
generalize this technique to the hardcore Bose-Hubbard model on a superlattice.

To determine the phase boundary separating the incompressible MI phase from the 
compressible SF phase within the SCE method, one needs the energy of the MI phase 
and its `defect' states -- those states which have one flipped spin (equivalently, one 
excited particle) about the ground-state -- as a function of the parameter $t$. At the point 
where the energy of the incompressible state becomes equal to its defect state, the system 
becomes compressible, assuming that the compressibility approaches zero continuously at 
the phase boundary. Note that these arguments are very similar to those presented in 
Sec.\ \ref{sec:ana} where exact results were obtained for the SF-vacuum insulators 
boundaries. Here however, the state of the system inside the MI phase is not known except for 
the special case $t=0$, where:
\beq
| \textrm{MI}^{(0)} \rangle &= 
|\uparrow\downarrow\uparrow\downarrow\ldots\uparrow\downarrow\uparrow\downarrow\rangle\,,
\eeq
where all the spin-ups (spin-downs) belong to the even (odd) sublattice. 

The energy of the MI phase is calculated via a many-body version of the nondegenerate 
Rayleigh-Schr\"odinger perturbation theory up to fourth order in $t$. We note that all 
odd-order terms in $t$ vanish for the $d$-dimensional hypercubic lattices considered in 
this manuscript. This is because this state cannot be connected to itself by only one 
hopping, but rather requires two hoppings to be connected.

Calculation of the wave functions and energies for the defect states is more involved as it 
requires the use of the many-body version of the degenerate Rayleigh-Schr\"odinger perturbation 
theory. The reason for that lies in the fact that when exactly one extra particle is added to 
the MI phase, it could go to any of the $N/2$ lattice sites that belong to the odd 
sublattice, since all of those states share the same energy when $t = 0$ (recall that $N$ 
is the number of lattice sites). Therefore, the initial degeneracy of the MI defect state 
is of order $N/2$. This degeneracy is lifted at second order in $t$, 
since all of the defect states occupy one of the sublattices, and they cannot be connected 
by one hopping, but rather require two hoppings to be connected. The wave function 
(to zeroth order in $t$) of the particle-defect state turns out to be
\beq
| \textrm{MI}^{(0)}_{\rm def} \rangle = \sum_{i \in \os} f_i \hat{S}_i^{+} |\textrm{MI}^{(0)}\rangle,
\eeq
where $f_i$ is the eigenvector of the matrix $T_{i i'}=\sum_{j \in \es} t_{ij} t_{ji'}$ 
with the highest eigenvalue, such that $\sum_{i' \in \os} T_{ii'} f_{i'} = 4 d^2 t^2 f_i$. 
Here, $t_{ij}=t$ for $\langle ij \rangle$ and zero otherwise. The normalization condition 
requires that $\sum_{i \in \os} |f_i|^2 = 1$. The eigenvector with 
the highest eigenvalue corresponds to the lowest energy state, i.e., to the ground state.
We calculate the energy of the $| \textrm{MI}^{(0)}_{\rm def} \rangle$ phase via degenerate 
perturbation theory up to fourth order in $t$. Here too all odd-order terms in $t$ vanish.

A lengthy but straightforward calculation leads to the following expression for the SF-MI 
boundary (for further details regarding the calculation, we refer the reader to a similar 
calculation given in Ref.\ \onlinecite{iskin09})
\beq
\label{eq:muMI}
\frac{\mu}{A} = 1 - \frac{d-1}{2d} x^2 - \frac{(d-1)(d-3)}{8 d^2} x^4 + O(x^6),
\eeq
where $x = 2dt/A$. This expression is exact for all $d$-dimensional hypercubic lattices up 
to the given order. In one dimension, it agrees with the analytical solution\cite{Val} of 
the model given by $\mu/A = 1$ (see Sec.\ \ref{sec:ibh}). In the $d \to \infty$ limit,
where the exact result is given by the mean-field expression, i.e.,  $\mu/A=\sqrt{1-x^2}$, 
Eq.\ (\ref{eq:muMI}) is the correct power-series expansion about $x = 0$.

In the two- and three-dimensional cases, fourth-order SCE is not very accurate near the 
tip of the MI lobe, as the variable $x$ is not very small there. Therefore, an extrapolation 
technique is desirable in order to determine the phase boundary more accurately. Such an 
extrapolation is possible for the MI phase, since it is already known for $d > 1$ that the 
critical point at the tip of the MI lobe has the scaling behavior  of a ($d$+1) $XY$ model. 
Therefore, we propose the following ansatz for the MI lobe which includes the known power-law 
critical behavior of the tip of the lobe:
\beq
\label{eq:smu}
\frac{\mu}{A} &=&\alpha_0 \left(1 + \alpha_1 x + \alpha_2 x^2 +
\alpha_3 x^3 + \alpha_4 x^4 \right) \nonumber \\&\times& 
(x_c - x)^{z\nu},
\eeq
where $x_c = 2dt/A_c$ is the critical point which determines the location of the MI lobe 
tip, and $z\nu$ is the critical exponent for the ($d$+1) $XY$ model which determines the 
shape of the MI lobe near $x_c$. The parameters $\alpha_i$ are determined by matching 
Eq.\ (\ref{eq:muMI}) with Eq.\ (\ref{eq:smu}), after the latter is expanded out to fourth 
order in $t$. This procedure leads to: 
\begin{subequations}
\begin{align}
\alpha_0 &= \frac{1}{x_c^{z\nu}} \,,\\
\alpha_1 &= \frac{z\nu}{x_c} \,,\\
\alpha_2 &= \frac{z\nu(z\nu+1)}{2x_c^2} + e_2 \,,\\
\alpha_3  &= \frac{z\nu(z\nu+1)(z\nu+2)}{6x_c^3} + \frac{z\nu}{x_c} e_2 \,,\\
\alpha_4 &= \frac{z\nu(z\nu+1)(z\nu+2)(z\nu+3)}{24x_c^4} \nonumber\\
&+ \frac{z\nu(z\nu+1)}{2x_c^2}e_2 + e_4 \,,
\end{align}
\end{subequations}
where $e_2 = -(d-1)/(2d)$ and $e_4 =-(d-1)(d-3)/(8 d^2)$ are the coefficients 
of the second and fourth order terms in our SCE.

In our extrapolations, we set $z\nu \approx 0.672$ for $d = 2$ and $z\nu = 1/2$ for $d > 2$. 
This leaves only $x_c$ to be fixed; something which is accomplished by a straightforward 
$\chi^2$ curve-fitting to the numerical data obtained in Sec.\ \ref{sec:SSE}. The results are 
shown by the solid lines in Figs.\  \ref{dim2} and \ref{dim3} for two and three dimensions, 
respectively. As one can immediately see, the SCE results are very 
accurate and provide an analytic expression for the phase boundaries. 

Alternatively, we can estimate $x_c$ using the above approach without fitting it to the 
numerical data. We do so by finding the value of 
$x_c$ for which the fifth-order term in $x$ of Eq.\ (\ref{eq:smu}) vanishes. This gives 
$x_c \approx 1.53$ for $d = 3$ ($ \approx 6.7\%$ error), and $x_c \approx 1.076$ for 
$d \to \infty$ ($\approx 7.6\%$ error).

Before moving on to the next section, we note here that a similar application of the 
SCE for the SF-VH phase boundary, where
\beq
| \textrm{VH}^{(0)}_{\textrm{def}} \rangle = \sum_{i \in \textrm{o.s.}} f_i \hat{S}_i^{-} 
| \textrm{VH} \rangle
\eeq
is the wave function (to zeroth order in $t$) of the hole-defect state, leads to
\beq
\frac{\mu}{A} = 1 + \frac1{2} x^2 - \frac1{8} x^4 + O(x^6),
\eeq
in agreement with the exact result derived in Sec.~\ref{sec:ana}, i.e., 
$\mu/A=\sqrt{1+x^2}$, up to the given order. In addition, we perform a SCE in $A$, 
and find that the large $x$ behavior of the phase boundary is given by 
$\mu/A = x + O(1/x)$, which is also in agreement with the exact result.

\section{\label{sec:md} Momentum distribution}

Having discussed the phase diagram of the hardcore Bose-Hubbard model with a superlattice
in the previous sections, next we analyze the momentum distribution $n(\mathbf{k})$ 
of these bosons. This quantity can be directly probed in experiments with ultracold atomic 
gases via an absorption imaging during a short time-of-flight.\cite{folling,spielman0708}
Since it is trivial to show that $n_\mathrm{VH} (\mathbf{k}) = 1$ in the VH phase, 
we shall concentrate only on the momentum distribution of the bosons in the MI phase, 
$n_\mathrm{MI} (\mathbf{k})$, where we will compare our numerical quantum Monte Carlo 
results with those of two analytical approaches: the random-phase approximation 
(RPA) and the SCE method introduced in the previous section.

The RPA is a well-defined linear operation in which thermal averages of products of 
operators are replaced by the product of their thermal averages.\cite{haley} Since the 
fluctuations are not fully taken into account in this method, it becomes exact only 
for infinite-dimensional bosonic systems, recovering the mean-field theory. This 
method has been recently applied to the onsite,\cite{sengupta, menotti-rpa} and 
extended\cite{iskin09-2} Bose-Hubbard models, and its results showed good qualitative 
agreement with the experiments in the former case.\cite{folling, spielman0708}
Here we apply this method to our model (for further details regarding the calculation, 
we refer the reader to a similar calculation given in Ref.\ \onlinecite{iskin09-2}), 
and obtain
\beq
n_{\mathrm{MI-RPA}}(\varepsilon_{\mathbf{k}}) = \frac{1}{2} 
\sqrt{ \frac{A-\varepsilon_{\mathbf{k}}} {A+\varepsilon_{\mathbf{k}}} },
\label{eqn:nkrpa}
\eeq
where 
$
\varepsilon_{\mathbf{k}} = -2t\sum_{i = 1}^d \cos(k_i a)
$
is the energy dispersion of noninteracting particles. Since the RPA phase boundary is exactly 
the same as the mean-field one, and it gives a critical value for $x = 2dt/A$ that is 
much smaller than the true critical value in finite-dimensions, we compare our results 
with a rescaled $x$ value such that
\beq
n_{\mathrm{MI-RPA}}^{\textrm{scaled}}(\varepsilon_{\mathbf{k}}) = \frac{1}{2} 
\sqrt{ \frac{A x_c-\varepsilon_{\mathbf{k}}} {A x_c + \varepsilon_{\mathbf{k}}}},
\label{eqn:nkrpascaled}
\eeq
where $x_c = 2dt/A_c$ is the true critical point which determines the 
location of the MI lobe tip. We call this the scaled RPA momentum distribution
following Ref.\ \onlinecite{freericks09}.

To extend the RPA result to finite dimensions, we also calculate 
$n_{\mathrm{MI}}(\varepsilon_{\mathbf{k}})$ as a power series expansion in the 
hopping $t$ via the strong-coupling perturbation theory. To second-order in $t$, 
we obtain (for further details regarding the calculation, we again refer the reader 
to a similar calculation given in Ref.\ \onlinecite{iskin09-2})
\beq
n_{\mathrm{MI-SCE}}(\varepsilon_{\mathbf{k}}) = \frac{1}{2} - 
\frac{\varepsilon_{\mathbf{k}}}{2A} + 
\frac{\varepsilon_{\mathbf{k}}^2 - 2 d t^2}{4A^2} + O(t^3),
\label{eqn:nksc}
\eeq
which is exact up to the given order for any dimension $d$. In the $d \to \infty$ limit (while 
$d t$ is kept fixed), we checked that Eq.~(\ref{eqn:nksc}) agrees with the RPA 
solution (which is exact in that limit) given in Eq.~(\ref{eqn:nkrpa}), when the latter is expanded out to second 
order in $t$. This provides an independent check of the algebra. 

\begin{figure*}[htp!]
\includegraphics[angle=0,scale=1,width=0.9\textwidth]{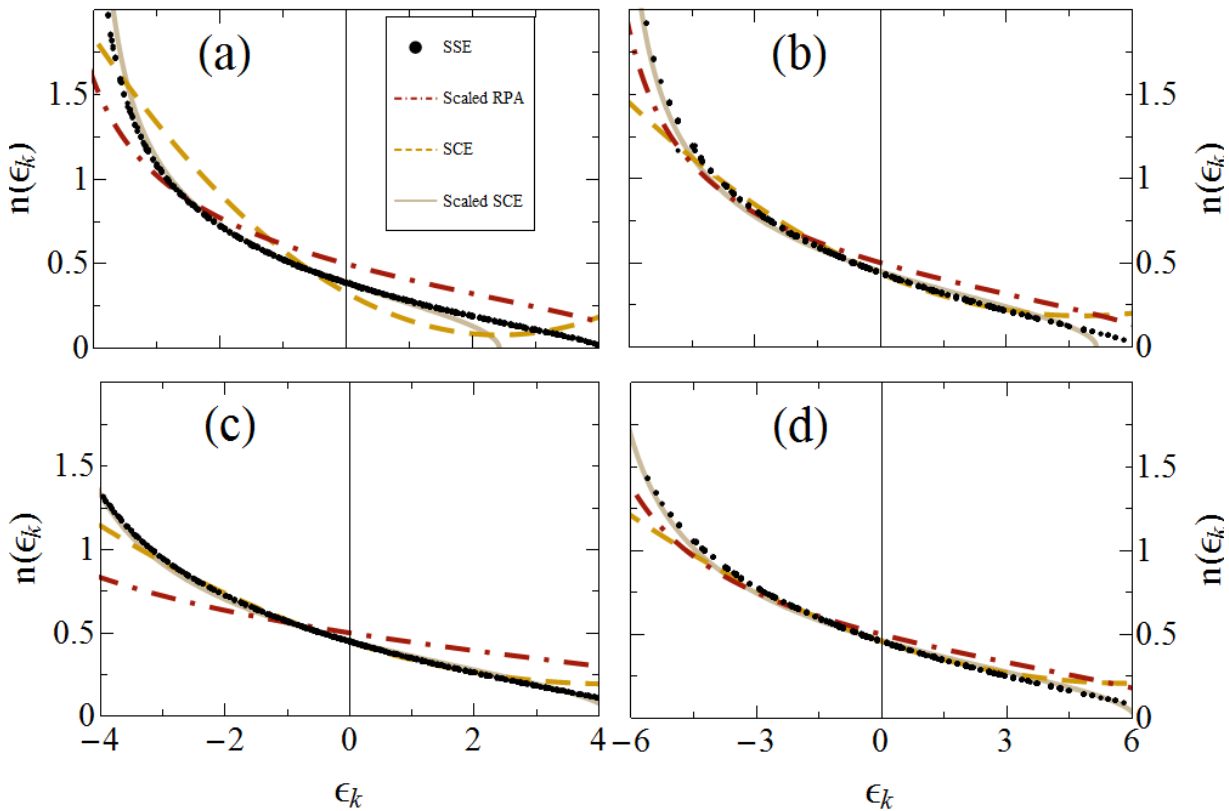}
\caption{\label{fig:mdf} (Color online) Momentum distribution function $n(\varepsilon_k)$ 
for two $48 \times 48$ systems: (a) $A=2.42 \quad (x \approx 1.653)$ and (c) $A=4.22 \quad (x \approx 0.948)$,
and two $14 \times 14 \times 14$ 
systems: (b) $A=4.8 \quad (x=1.25)$ and (d) $A=6 \quad (x =1)$. The full circles are the numerical SSE results. The scaled 
RPA is the dot-dashed line, whereas the dashed and solid lines are the SCE and scaled SCE, 
respectively. The figures show that the scaled SCE results are much better than any of the 
other two approximation methods, and that the scaled SCE results fit better,
as $A$ becomes larger ($t=1$ in all four systems) -- suggesting we are deeper inside the MI phase.
}
\end{figure*}

The second-order SCE is not very accurate near the tip of the MI lobe, as $t/A$ is not small 
there. To extend its region of validity, we therefore propose the following ansatz, 
\beq
n_{\mathrm{MI}}(\varepsilon_{\mathbf{k}}) = \frac{1}{2} 
\sqrt{ \frac{A - \varepsilon_{\mathbf{k}} + 
(4\lambda-2)dt^2/A} {A + \varepsilon_{\mathbf{k}} + 4\lambda dt^2/A} } \,
\label{eqn:nkansatz}
\eeq
for any dimension $d$, where $\lambda = d(x_c-1)/x_c^2$ depends on $d$. This expression reduces 
to Eq.~(\ref{eqn:nkrpa}) in the $d \to \infty$ limit, and it has the 
correct power-series expansion about $x = 0$ up to second-order in $t$, 
i.e., Eq.~(\ref{eqn:nksc}). We call this the scaled SCE momentum distribution.

In Fig.\ \ref{fig:mdf}, we show several comparisons (two in two dimensions and two in three 
dimensions) between the momentum distribution function obtained with the quantum Monte Carlo 
and the three approximations obtained above, namely, the scaled RPA, the SCE, and the scaled 
SCE. As the figures indicate, the scaled SCE is a far better fit than the other two methods, 
and more so for larger values of $A$, that is, deeper inside the MI phase where the SCE becomes more and more accurate. 

\section{\label{sec:conc}Conclusions}

We have obtained the complete phase diagram of the hardcore Bose-Hubbard model with a 
period-two superlattice in two and three dimensions. First we have calculated 
the boundaries between the superfluid phase and the `trivial' insulators (the completely-empty 
and completely-filled lattices) analytically. Then, using quantum Monte Carlo simulations followed 
by a finite-size scaling, we have determined the phase boundary between the superfluid 
phase and the (half-filled) Mott insulator. We have also compared our numerical 
results against three approximation schemes: the usual Gutzwiller mean-field approach,
a cluster mean-field approach, and the strong-coupling expansion (SCE) method. 

For the transition between the superfluid phase and the `trivial' completely-empty 
and completely-filled lattice insulators, we have found that the mean-field approaches 
yield the exact results in any dimension. As for the superfluid-Mott insulator 
boundary, the Gutzwiller approach was shown to work very 
poorly (up to $\approx 100\%$ error in two dimensions and $\approx 50\%$ error in three 
dimensions). This is a clear indication of the fact that this mean-field approach
is not suitable for describing the superfluid-Mott insulator transition in the vicinity 
of the tip of the lobe. A cluster mean-field approximation scheme, which is based on the 
underlying checkerboard symmetry of the problem, was proven to be a big improvement over the 
previous method (reducing the error to one half of the one generated by the usual 
Gutzwiller ansatz), albeit still far from being accurate as one approaches the tip of 
the Mott lobe. 
The fourth-order SCE turned out to be the best method among the three in describing the 
superfluid-Mott insulator phase boundary, as the one-parametric fit of the SCE yielded 
very accurate results, also near the tip of the Mott lobe where the other methods failed. 
It also provided an analytic expression for that boundary, which could be used as a guide 
in future experimental realizations of this model. 

Finally we have examined the extent to which several approximation schemes, 
such as the random phase approximation and the strong-coupling expansion, give an 
accurate description of the momentum distribution of the bosons inside the 
insulating phases. We have shown that a scaled SCE provides an accurate
analytic expression for the momentum distribution of the bosons	 inside the Mott-insulating 
phase both in two and three dimensions, which could again be used as a guide 
in future experimental realizations of this model.

\begin{acknowledgments}
I.H and M.R. were supported by the US Office of Naval Research under Award No.\ N000140910966,
and M.I. thanks The Scientific and Technological Research Council of Turkey 
(T\"{U}B$\dot{\mathrm{I}}$TAK) for financial support.
We are grateful to J. K. Freericks and H. R. Krishnamurthy for useful discussions.
\end{acknowledgments}

\end{document}